\documentclass[10pt,conference]{IEEEtran}
\IEEEoverridecommandlockouts
\usepackage{cite}
\usepackage{amsmath,amssymb,amsfonts}
\usepackage{algorithmic}
\usepackage{graphicx}
\usepackage{textcomp}
\usepackage{xcolor}
\usepackage{url}
\usepackage{hyperref}
\def\BibTeX{{\rm B\kern-.05em{\sc i\kern-.025em b}\kern-.08em
    T\kern-.1667em\lower.7ex\hbox{E}\kern-.125emX}}

\newcommand{\bpstart}[1]{\vspace{1mm} \noindent{\textbf{#1.}}}

\begin{document}

\title{Interaction Techniques for User-friendly Interfaces for Gate-based Quantum Computing}

\author{\IEEEauthorblockN{Hyeok Kim}
\IEEEauthorblockA{\textit{Computer Science} \\
\textit{Northwestern University}\\
Evanston, U.S.A. \\
% 0000-0003-4340-4470
hyeok@northwestern.edu}
\and
\IEEEauthorblockN{Kaitlin N. Smith}
\IEEEauthorblockA{\textit{Computer Science} \\
\textit{Northwestern University}\\
Evanston, U.S.A. \\
% 0000-0002-1169-3696
kns@northwestern.edu}
}

\maketitle

% abstract: 200-250 words
\begin{abstract}
Quantum computers offer promising approaches to various fields. 
To use current noisy quantum computers, developers need to examine the compilation of a logical circuit, the status of available hardware, and noises in results.
As those tasks are less common in classical computing, quantum developers may not be familiar with performing them.
Therefore, easier and more intuitive interfaces are necessary to make quantum computers more approachable.
While existing notebook-based toolkits like Qiskit offer application programming interfaces and visualization techniques, it is still difficult to navigate the vast space of quantum program design and hardware status.

Inspired by human-computer interaction (HCI) work in data science and visualization, our work introduces four user interaction techniques that can augment existing notebook-based toolkits for gate-based quantum computing: (1) a circuit writer that lets users provide high-level information about a circuit and generates a code snippet to build it; (2) a machine explorer that provides detailed properties and configurations of a hardware with a code to load selected information; (3) a circuit viewer that allows for comparing logical circuit, compiled circuit, and hardware configurations; and (4) a visualization for adjusting measurement outcomes with hardware error rates.
\end{abstract}


\section{Introduction}
With enhanced computing power, quantum computers enable a wide range of problems that classical computers have not been able to solve efficiently~\cite{b1}, such as drug discovery, physics, and cryptography.
To make the best use of noisy intermediate-scale quantum (NISQ) era quantum computers, users need to understand properties of a hardware device to use, such as gate errors and qubit coherence time (T1/T2). 
To encourage large-scale adoption and widely support various domain areas, quantum computers will need to equip easy and intuitive user interfaces for domain experts with varying levels of expertise.
Furthermore, research on quantum computing has been working on algorithms and architectures for small-to-middle scale circuits to leap from the current state to future generations like fault-tolerance.
Easy-to-use and well-streamlined user interfaces will also help those researchers to best use of their expertise.

\begin{figure*}[!htbp]
    % \centering
    \centerline{\includegraphics[width=0.95\textwidth]{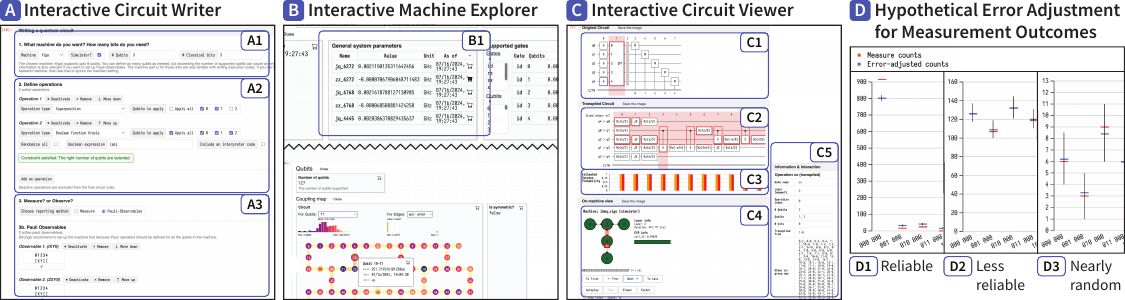}}
    \caption{Prototypes interfaces on a Jupyter Notebook environment with Qiskit~\cite{b2}.}
    \label{fig:main-figure}
\end{figure*}

While we are fortunate to have notebook-based libraries like Qiskit~\cite{b2}, Bloqade~\cite{b3}, and Strawberry Fields~\cite{b4}, for example, they often lack intuitive interfaces.
While these toolkits are built to support common tasks like writing and optimizing a circuit, finding a hardware, and analyzing the results,
they often provide application programming interfaces with basic documentations.
Thus, navigating the vast space of circuit design relies more on individual developers' tedious efforts, making it challenging for non-experts to map their real-world problems to the building blocks of quantum logic. 
While web-based tools provide relevant information visually, they are often disconnected from working environments, causing programmers to switch working contexts.

To improve quantum computing toolkits' interfaces, human-computer interaction (HCI) research on data science and visualization provide an inspiring direction given the similarity between quantum computing pipelines and common data science work pipelines that involve statistical modeling, model refinement, and result analysis~\cite{b8}.
For example, tools like Tisane~\cite{b9} and EVM~\cite{b10} let data scientists interactively build and revise model with less coding, allowing both novice and expert data scientists to easily perform statistical modeling.

Inspired by such approaches, our work proposes four (user) interaction techniques as a first step toward building more user-friendly interfaces for gate-based quantum computing, accompanied by demonstration prototypes (as extensions to Qiskit; \url{https://see-mike-out.github.io/hqci-demo/}).
First, we introduce an interactive circuit writer that allows build a quantum circuit with less experiences in quantum coding. 
Second, we propose an interactive machine explorer that shows properties of a quantum computer and generates a reusable code snippet for selected properties.
Third, we show an interactive visualization that connects a logical circuit, its optimized circuit, and how a machine operates it with fidelity information.
Lastly, we suggest a visualization for \textit{hypothetical error-adjusted measurement outcomes} via a Monte-Carlo approach.

\section{{Interface Techniques for Quantum Computing}}
As depicted in Figure~\ref{fig:main-figure}, we propose four interaction techniques for gate-based quantum computing interfaces.

\bpstart{(1) Circuit Writer}
Common hurdles for writing a quantum circuit include converting conceptual ideas into operations (e.g., superposition to a Hadamard gate) and validating those operations.
An interactive circuit writer (Fig.~\ref{fig:main-figure}A) could benefit those with limited experiences in low-level circuit design by mediating a conceptual circuit to codes.
This interface allows a developer to identify a qubit register (A1), add conceptual operations (A2), and set measurement or Pauli observables (A3). 
For example, when a developer chooses oracle-based Grover's search, the interface lets them write an boolean expression for the oracle instead of low-level gates. 
The interface validates the circuit (e.g., the number of qubits, observable expressions) and shows warnings for any invalid part.
The developer can retrieve the Qiskit code for the circuit. 

\bpstart{(2) Machine Explorer}
As current quantum machines are improving on stability, developers need to check the status of a machine to use. 
Yet, finding relevant information often needs moving back and forth between the development environment, documentation, and the machine's platform website. 
To reduce this working context switch, we introduce an interactive machine explorer (Fig.~\ref{fig:main-figure}B) that shows a machine's status and properties in a dashboard.
A developer can select interesting or important properties (B1), then the interface generates a reusable code snippet to allow for easily accessing the same properties next time.
We expect that the interactive machine explorer could assist in algorithm-machine pairing.

\bpstart{(3) Circuit Viewer}
While a logical circuit can include any gates, 
it has to be decomposed into basic gates supported by a machine.
For developers who work on improving such compilation process or want to understand it, we introduce an interactive circuit viewer that links a logical circuit (C1) and its compiled circuit (C2).
When a developer selects a gate in the compiled circuit, then the corresponding logical gate is highlighted (and vice versa). 
Each layer in the compiled circuit is also mapped to layer-wise and cumulative Estimated Success Probabilities (ESPs)~\cite{b11} (C3).
For Layer $i$, the layer-wise ESP is the product of the success rates ($= 1 - $ error) of the layer's gates, and the cumulative ESP is the product of the success rates of the gates from Layer 0 to $i$.
In addition, it provides an on-machine view animation (C4) that shows how a machine operates each layer with cumulative ESPs mapped to the qubits' color. 
Lastly, an on-demand detail information panel is provided for a selected visual element (C5).

\bpstart{(4) Hypothetical Error Adjustment for Measurement Outcomes}
As current quantum computers exhibit non-negligible errors, a developer must consider operation errors in analyzing results and determining the number of shots.
To support that, we propose a visualization for \textit{hypothetically error-adjusted measurement outcomes} using an Monte-Carlo approach~\cite{b12}. 

At high-level, we propose a procedure that simulates whether gates and measurements are erroneous using their error rates.
We then maps each measurement outcome to the same value if the simulation procedure returns no error; or something else (randomly chosen) otherwise.
By repeating this simulation a lot of times (10,000+), we obtain a distribution of error-adjusted measurement outcome counts (blue ticks).
For instance, too few shots (D3) or a wrong circuit (D2) exhibits overlapping 95\% confidence intervals (error bars) while a valid circuit with enough shots do not (D1). 

\section{Future Work and Conclusion}
Our work introduces several approaches to making quantum computing more accessible via interface techniques, including circuit writer, machine explorer, circuit viewer, and uncertainty visualization, inspired by prior HCI work on data science and visualization. 
We demonstrate our approaches using a prototype demo interface with Qiskit (\url{https://see-mike-out.github.io/hqci-demo/}).
Interesting future research includes user study-based validation of these techniques and the productization of them with enhanced scalability.

\end{document}